\documentclass{llncs}

\usepackage{amssymb}
\usepackage{graphicx}
\usepackage{xspace}
\usepackage{url}
\usepackage[draft]{fixme}
\urldef{\mailsa}\path|{wachs, herold, posselt, dold, carle}@net.in.tum.de|
\newcommand{\keywords}[1]{\par\addvspace\baselineskip
\noindent\keywordname\enspace\ignorespaces#1}

\usepackage[disable]{todonotes}
\newcommand{\gplmt}{GPLMT\xspace}
\newcommand{\pl}{PlanetLab\xspace}

\usepackage{fancyvrb}
\usepackage{multicol}
\usepackage{pgfkeys}
\usepackage{listings}
\usepackage{color}

\begin{document}

\mainmatter  

\title{\gplmt: A Lightweight Experimentation and Testbed Management Framework}

\author{Matthias Wachs \and Nadine Herold \and Stephan-A. Posselt \and Florian
Dold \and Georg Carle}

\institute{Technical University of Munich (TUM),\\
Boltzmannstr. 3, 85748 Garching bei M\"unchen, Germany\\
\mailsa\\
\url{https://www.net.in.tum.de}}

\maketitle

\begin{center}
\textbf{This is a preprint of the publication to appear at PAM 2016. \newline
Last update: 12 Jan 2016.}
\end{center}

\begin{abstract}
Conducting experiments in federated, distributed, and heterogeneous testbeds is
a challenging task for researchers. Researchers have to take care of the whole
experiment life cycle, ensure the reproducibility of each run, and the
comparability of the results. We present \gplmt, a flexible and lightweight
framework for managing testbeds and the experiment life cycle. \gplmt provides
an intuitive way to formalize experiments. The resulting experiment description
is portable across varying experimentation platforms. \gplmt enables
researchers to manage and control networked testbeds and resources, and conduct
experiments on large-scale, heterogeneous, and distributed testbeds. We state
the requirements and the design of \gplmt, describe the challenges of
developing and using such a tool, and present selected user studies along with
their experience of using \gplmt in varying scenarios.

\end{abstract}

\section{Introduction}

Network testbeds are an invaluable tool for researchers developing network
protocols and networked systems. Local testing and simulation can be a first
step to prove the viability of an approach, but has the drawback of abstracting
and neglecting important properties of the real world. Testing and deploying a
novel approach in software which is actually deployed ``in-the-wild'' is the
next step to analyze the impact and behavior on real-world networks.

A testbed may be heterogeneous with respect to the hardware and the
operating system, and may be physically distributed across more than one
location. This allows the researcher to evaluate reliability and portability
under close to real-world conditions. However, an experiment is challenging to
manage in a complex testbed. The life cycle of a network experiment starts with
tasks such as testbed configuration, resource allocation, experiment
definition, and deployment. The execution plan may require assigning different
tasks to subsets of nodes in a precise timely manner to control the execution.
At the end, the results need to be collected from all nodes. Monitoring and
error handling also have to be considered, as resources may become unavailable,
or a sub-task may fail. At worst, an experiment lasting several days
has to be repeated.

A large variety of testbeds is available to researchers. Many of them focus on
a specific domain (e.g. wireless experimentation, high precision measurements,
real-world network testbeds), and most of them use a proprietary and
domain-specific approach to how the testbed is designed, accessed, managed, and
experiments are controlled, requiring a manual adaptation for every experiment.
When trying to transfer such an experiment to a different testbed, the
experimenter has to adapt---and most of the time rewrite---the experiment to be
able to transfer the experiment to a different platform. This makes it difficult to
reproduce and confirm experiment results for both the researcher as well as the
research community.

These tasks are similar to many experiments but are still performed by most
experimenters manually, or with the help of ad-hoc scripts which are rarely
reusable. Instead of implementing ad hoc solutions specific to our particular
problems, we decided to realize a flexible and extensible testbed and
experimentation tool, supporting us in our work and to make it available to the
public.

With this work, we present \gplmt, a flexible, lightweight experimentation and
testbed management tool. \gplmt provides an intuitive way for users to define
experiments, supports the full experimentation life cycle, and allows experiments
to be transferred between different testbeds and platforms, ensuring
reproducibility and comparability of experiment results. \gplmt is free software
and its source code is publicly available on the \gplmt
website\footnote{\url{https://github.com/docmalloc/gplmt}}. In the remainder of
this paper we will give an overview of \gplmt, state the requirements and
challenges for such a tool, and describe the design and implementation. In
Section~\ref{sec:usecase}, we describe the experiences of users working with
\gplmt in various scenarios.

\section{\gplmt Features}
\label{sec:features}

\gplmt is started on a control node and executes a user-supplied XML-based
experiment description. \gplmt provides an experiment definition language to
define the resources participating in the experiment, the tasks to execute and
including specific order and parallelism, and to assign such tasks to
resources.  In addition, it allows the inclusion of files to reuse experiment
definitions and to group resources.  \gplmt supports the \pl-API and supports
importing information about available and assigned nodes from the user's \pl
account.

\gplmt connects to the nodes via an arbitrary communication channel (such as SSH),
runs tasks on the nodes, i.e. platform-specific binaries or executable scripts,
and can transfer files between the controller and the nodes. \gplmt offers additional features focusing on handling the intricacies of
testbeds: the user can annotate commands with different modes of
failure and register arbitrary cleanup actions to, for example, kill processes and
delete temporary files.

\section{Related Work}

Various different tools exist to manage and control network experiments. A
rather extensive list can be found on the \pl
website\footnote{\url{https://www.planet-lab.org/tools}}. \cite{2015jaros}
provides a comprehensive analysis with respect to quality and usability of such tools,
finding most of them not usable or suitable to be used with respect to today's
network experiments. Many of these tools are outdated and not available anymore
(Plush, Nebula, Plman, AppManager) or were not even made publicly available at
all (PLACS). Some of these tools provide rather basic functionality to invoke
commands on remote nodes (pssh, pshell, vxargs) not supporting error conditions
and error handling as well as orchestrating nodes to perform complex and synchronized
operations. The Stork project\footnote{\url{http://www.cs.arizona.edu/stork/}}
provides a deployment tool for \pl nodes including configuration. This tool
lacks fine-grained execution control to setup more complex experiments.
Gush (GENI User Shell)~\cite{gush} claims to be an execution management
framework for the GENI testbed. Gush provides extensive methods to define
resources but is limited regarding control flow aspects. Parallel or sequential
execution is not possible in a straight forward manner. In addition, Gush is
not longer supported\footnote{\url{http://gush.cs.williams.edu/trac/gush}}.

Experimentation frameworks like NEPI~\cite{6064394} require the user to do
rather complex adaptations in the source code to extend it with new
functionalities and add support for new platforms. Approaches like
OMF~\cite{Rakotoarivelo_OJS_10} focus on the management and operation of network
testbed infrastructures and federation between infrastructures not focusing on
the experiment part in the life cycle.

The COCOMA framework~\cite{ragusa} focuses on providing an experimentation
framework for cloud based services to control and execute tests for cloud based
services in a controlled and reproducible manner and to study resource
consumption of such services. \cite{NICS} proposes an emulated testbed for the
domain of cyber-physical systems. This work focuses more on the testbed
implementation and less on the execution of experiments.

\section{Requirements and Challenges}\label{requirements}

In this section, we highlight the requirements for the design of an
experimentation and management tool realizing the features described in
section~\ref{sec:features} and based on experiences obtained from conducting
different types of experiments with various testbeds, exchange with the research
community and an analysis of possible use cases varying from managing large
scale and unreliable to small virtualization based testbeds.

\textbf{Self-Containment.}
\gplmt is intended as a lightweight tool for researchers and experimenters. The
tool should neither require a complex experimentation infrastructure, rely on
client software like agents installed on testbed nodes nor have requirements for
external services like a database server. The tool shall be realized
as a portable, platform independent stand-alone tool.

\textbf{Scalability }
is important for the experimentation tool to support large-scale
testing and experimentation. When conducting experiments with many participants,
orchestration and controlling of a large number of different nodes is a
challenging task since large delays and setup times have to be prevented.

\textbf{Resource Restrictions.}
Experimentation with \gplmt may be limited due to restrictions in the
surrounding environment.  Establishing a large number of connections to a large
number of nodes has to be realized efficiently. Therefore, \gplmt has to be aware
of resource restrictions in the host environment and reuse
connections and provide rate limiting for new connections being established.

\textbf{Heterogeneous Testbeds and Nodes.} \gplmt has to make experimentation
independent from the testbed platform and the participating nodes.  Experiments
have to be executable in heterogeneous environments with different operating
systems and different versions of the operating system.

\textbf{Fault Tolerance in Unreliable Environments.} In real-world and large-scale
network testbeds availability of resources cannot always ensured: not all
assigned nodes and resources may be available or can fail during an experiment
and become available again. \gplmt, therefore, has to cope with unreliable
resources and has to provide automatic error handling and recovery transparent
to the experiment.

\textbf{High-level Experiment Definition.}  With \gplmt experiment definition 
shall be done on a high level of abstraction, to allow the experimenter to focus 
on essential aspects of experiment design and control flow without getting
distracted by implementation details.

\textbf{Experiment Reproducibility.} Experiment reproducibility is essential for
confirmability of experimental results. \gplmt has to support an experiment flow
making execution independent from participants, resources, testbeds, external
dependencies and state based on a high-level definition of experiments.

\textbf{Experiment Portability, Reusability and Extensibility.} Experiments shall be
transferable to other testbeds infrastructures and allow researchers to share
experiment definitions. Employing an abstraction over the testbed infrastructure
and using a high-level description of an experiment allows an experiment
definition to be reused and to be varied in different scenarios speeding up the
testing process.


\textbf{Grouping Entities in Experiments.} In an experiment, tasks and resources
may be assigned to different groups of nodes. \gplmt shall provide the
functionality to group nodes and resources and to assign tasks to such a group.

\textbf{Nested Task Execution and Synchronization.} Within an experiment, tasks
often have to be executed in a specific order or can be executed in parallel.
\gplmt shall provide constructs to allow experimenters to specify the execution order of
tasks. Tasks may also be nested and grouped in such
sequential and parallel constructs. Additional synchronization barriers between
the tasks have to be provided.

\textbf{Repeatable, Periodic and Scheduled Tasks for Experiments.}
Often tasks inside an experiment have to be executed repeatedly or
triggered periodically or at a certain point in time (e.g. for periodic
measurements). \gplmt has to provide constructs to express a looping
functionality and to schedule tasks to be
executed at certain point in time or after a certain duration without adding high complexity.

\textbf{Error Condition Handling in Experiments.}
In many cases the experiment control flow depends on successful or failed
execution of tasks, making subsequent operations useless or
the whole experiment fail. Therefore, \gplmt has to allow the experimenter to
define the expected result of a task and how an error condition has to be
handled.
In addition, functionality to define a clean up and tear down
task---executed before the experiment is terminated---is beneficial.

\section{\gplmt Design and Implementation}

In this section we present the design and implementation of \gplmt, which were
driven by the requirements described in Section~\ref{requirements}.

\subsection{Architecture}

\gplmt is designed as a stand-alone tool running on the so-called \emph{\gplmt
controller}. The \gplmt controller is responsible for orchestrating the whole
experiment, i.e. scheduling tasks on the hosts of a testbed, from now on called
\emph{nodes}. \gplmt manages a connection from the controller to each node.
\gplmt does not require any original services on the nodes, but relies on SSH,
and possibly other protocols in the future. In addition, \gplmt can use the
\pl-API to obtain information about available nodes in the experimenter's \pl
slice.

An experiment is conducted by passing an experiment description in a high-level
description language to \gplmt. The description tells \gplmt which nodes to
connect to, which files to exchange, and which tasks to run.


\subsection{Resource Management}

In large-scale experiments with many nodes, \gplmt will open a large number of
connections. SSH is particularly resource-intense. The SSH connection setup is
computationally expensive due to cryptography and may overload a low-powered
controller or the physical host of a virtualized testbed. A high rate of
connection attempts may stress IDS systems, and may trigger IDS alerts for
alleged SSH scanning.

\gplmt offers two solutions to limit its resource usage: \emph{connection
reuse} and \emph{rate limiting} of connection attempts. \gplmt will tunnel all
commands to the same node through a single control connection, but will still
try to reconnect when the connection is lost. \gplmt optionally delays
connection attempts, including reconnects, to not exceed a configurable number
of attempts per interval.

%

%
%
%

\subsection{Implementation}

The \gplmt controller is implemented in Python 3. Besides a few Python
libraries and the Python interpreter itself, \gplmt only depends on the
external tools which are needed to connect to nodes. Notably, \gplmt wraps
OpenSSH, so all features of OpenSSH are available via a local OpenSSH
configuration file. \gplmt directly uses OpenSSH's \emph{control master}
feature to reuse connections to the same node.

\section{\gplmt's Experiment Definition Language}
\label{sec:language}

\gplmt provides a domain-specific language to describe the experiment setup and
execution. Its syntax is defined in an \textit{XML Schema} obtained from a
\textit{relax-ng} definition. Therefore, terms such as \textit{element} and
\textit{attribute} refer to the respective XML objects.

The \texttt{experiment} root element may contain multiple \texttt{include},
\texttt{targets}, and \texttt{tasklist} elements and a single \texttt{steps}
element. A \texttt{targets} element names the nodes and can also be used to
group nodes. \texttt{tasklist} defines a set of commands to be run.  Both
definitions are tied together with the \texttt{steps} element, which states
which tasklist is to be executed on which targets and at what time.

Target and tasklist definitions are optional and may also be imported from
other documents. Targets and tasklists are distinguished and referenced by
unique names.

\subsection{Targets}
\label{targets}

A \texttt{target} element names a member node, and specifies how to
access the node. The following types of targets are currently supported:
\begin{itemize}
	\item \texttt{local} specifies execution on the \gplmt controller itself.
	\item \texttt{ssh} states that the nodes can be accessed using SSH. The
		child elements \texttt{username} and \texttt{password} may provide
		credentials.
	\item \texttt{planetlab} specifies a \pl node and accepts the \pl-API-URL,
		the slice, and the user name as attributes.
	\item \texttt{group} specifies a nested target definition, creating a set
		of nodes (and other groups) addressable as a single target.
\end{itemize}

To support parameterization per target, each target
definition can contain multiple \texttt{export-env} elements, which declare an
environment variable to be exported. The value of this variable is then
available to tasks on the target.

\subsection{Tasklists}
\label{tasklist}

The \texttt{tasklist} binds a list of \textit{tasks} to a name. A \textit{task}
is one of the following predefined commands:
\begin{itemize}
	\item \texttt{get} and \texttt{put} are used to exchange files between the
		controller and the targets.
	\item \texttt{run} accepts a command to be executed.  When a
		target defines additional environment variables, those are passed to
		the command using \texttt{export-env}.
	\item The \texttt{par} and \texttt{seq} elements contain nested lists of
		tasks.  \texttt{seq} will run those tasks in order, whereas
		\texttt{par} will immediately start all sub-tasks in parallel.
	\item \texttt{call} is used to reference a tasklist to be executed.
\end{itemize}

\noindent \texttt{tasklist} accepts the optional attributes \texttt{cleanup},
\texttt{timeout}, and \texttt{error}, controlling the tasklist's behavior in
case of an error condition. \texttt{cleanup} references another tasklist to be
executed after the current tasklist, even if the current tasklist aborts due to
an error. This can be used to kill stale processes and delete temporary files or to
save intermediate results. \texttt{timeout} specifies the maximum amount of time
the tasklist is allowed to execute before it is aborted. This guarantees
progress in case a command loops infinitely or dead-locks. \texttt{on-error}
determines how \gplmt continues when a task fails. The following fail modes are
available:

\begin{itemize}
	\item \texttt{abort-tasklist} aborts the current tasklist and
		continues with the tasklist specified by the surrounding context.
\todo{step not introduced}{	\item \texttt{abort-step} aborts the current \textit{step} and continues
		with the next \textit{step}. Steps are explained in Section~\ref{steps}.}
	\item \texttt{panic} aborts the whole experiment.
\end{itemize}

\subsection{Steps}
\label{steps}

The language requires exactly one \texttt{steps} element. It may contain
multiple \texttt{step}, \texttt{synchronize}, \texttt{register-teardown}, and
\texttt{repeat} elements.

The \texttt{step} element determines which tasklists run on which target. A
start and a stop time can be added to schedule a task for later execution. Times
are either relative to the start of the experiment or absolute wall clock times,
allowing to defer a step until night-time when resources are available. Thus,
\texttt{step} elements form the basic building block for orchestrating the
experiment.

Consecutive \texttt{step} elements run in parallel. A \texttt{synchronize}
element represents barrier synchronization, and execution can only continue
after all currently running steps have finished.

\texttt{register-teardown} references a tasklist by name that is executed when
\texttt{steps} finishes. This tasklist is always executed, even if errors lead
to the abortion of the experiment. The registered tasklist is intended to
contain cleanup tasks and to transfer experiment results to the controller. The
\texttt{register-teardown} cleanup tasklist only needs to be registered right
before the \texttt{step} that allocates the corresponding resources is issued.

\gplmt's experiment definition language offers basic loops within
\texttt{steps}: The \texttt{repeat} element loops over the enclosed steps until
at least one of the following conditions is satisfied:
\begin{itemize}
	\item a given number of iterations (\texttt{iterations})
	\item a given amount of time has passed (\texttt{during})
	\item a given point in time was passed (\texttt{until})
\end{itemize}
These are deliberately simple conditions that only allow for decidable loops,
so it can be easily verified by manual inspection (or programmatically) whether
a loop terminates.

\subsection{Example}

\begin{lstlisting}[language=XML,
 frame=lrtb,
 float=p,
 xleftmargin=6ex,
 label=lst:full-ex,
 tabsize=2,escapechar=\#,
 caption={Example: Generate and Monitor Network Traffic with \gplmt}]
<?xml version="1.0" encoding="utf-8" ?>
<experiment>

 <include file="include/teardowns.xml" />  #\label{l:include}#

 <targets>  #\label{l:targets}#
   <target name="monitor" type="ssh">  #\label{l:monitor}#
     <user>testaccount</user>
     <host>monitor.example</host>
   </target>
   <target name="pingGroup" type="group">  #\label{l:pingGroup}#
     <target name="A" type="ssh">
       <user>testaccount</user>
       <host>10.0.0.16</host>
       <export-env var="host" value="10.0.0.17" />  #\label{l:varA}#
    </target>
     <target name="B" type="ssh">
       <user>testaccount</user>
       <host>10.0.0.17</host>
       <export-env var="host" value="10.0.0.16" />  #\label{l:varB}#
     </target>
   </target>  #\label{l:/pingGroup}#
 </targets>  #\label{l:/targets}#

 <tasklists>  #\label{l:tasklists}#
   <tasklist name="createPCAP">  #\label{l:createPCAP}#
     <run>tcpdump -i eth0 -w testrun.pcap &</run>  #\label{l:tcpdump}#
   </tasklist>
   <tasklist name="doPing">  #\label{l:doPing}#
     <run>ping #\$#host -c 10</run>  #\label{l:ping}#
   </tasklist>  #\label{l:/doPing}#
   <tasklist name="getData">
     <get>testrun.pcap</get>
   </tasklist>
 </tasklists>  #\label{l:/tasklists}#

 <steps>  #\label{l:steps}#
   <step tasklist="createPCAP" targets="monitor" />  #\label{l:s-createPCAP}#
   <register-teardown ref="stopMonitoring"  #\label{l:s-stopMonitoring}#
     targets="monitor" />
   <synchronize />  #\label{l:sync-monitor}#
   <step tasklist="doPing" targets="pingGroup" />  #\label{l:s-doPing}#
   <synchronize />  #\label{l:sync-pings}#
   <step tasklist="getData" targets="monitor" />  #\label{l:getData}#
 </steps>  #\label{l:/steps}#
</experiment>
\end{lstlisting}

In this section, we present a brief example for a \gplmt experiment to
illustrate how experiments are defined. In this experiment, we use \gplmt,
running on the controller, to generate network traffic on two nodes and capture
this traffic using a third monitoring node. Therefore, nodes $A$ (IP 10.0.0.16)
and $B$ (IP 10.0.0.17) ping each other. The $monitor$ collects all network
traffic using \textit{tcpdump}.  At the end of the experiment, the resulting
capture file is transferred to the controller. Listing~\ref{lst:full-ex} shows
a (slightly abbreviated) description for this experiment.

First of all, an external experiment description containing teardown
functionality is included (l.~\ref{l:include}). Separating functionality in
different files eases reuse of frequently used targets and tasklists.

The definition for the three nodes $A$ and $B$ and $monitor$ is done in the
\texttt{targets} element (ll. \ref{l:targets}--\ref{l:/targets}): nodes $A$  and
$B$ are grouped into a target named \texttt{pingGroup}.
To ping each other, these hosts have to know the partner's IP address which
is provided in the environment variable \texttt{host}.

The experiment workflow is defined in the \texttt{steps} element (ll.
\ref{l:steps}--\ref{l:/steps}). The different \texttt{step} elements reference
tasklists from the \texttt{tasklists} element (ll.
\ref{l:tasklists}--\ref{l:/tasklists}). The experiment starts with instructing
the monitor node to capture network traffic using \texttt{tcpdump}  (l.
\ref{l:s-createPCAP}) using tasklist \texttt{createPCAP}
(l.~\ref{l:createPCAP}). To ensure \texttt{tcpdump} is terminated at the end of
the experiment, the experiment registers tasklist \texttt{stopMonitoring}
(l.~\ref{l:s-stopMonitoring}), imported from a file (l.~\ref{l:include}).
Both tasklists, \texttt{createPCAP} and \texttt{stopMonitoring}, are executed in
parallel.

The \texttt{synchronize} statement (l. \ref{l:sync-monitor}) ensures
monitoring is started before the nodes in group
\texttt{pingGroup} (ll. \ref{l:pingGroup}--\ref{l:/pingGroup}) begin
to ping each other (l. \ref{l:s-doPing}). Both nodes execute the same tasklist
\texttt{doPing} (ll. \ref{l:doPing}--\ref{l:/doPing}). The shell on
respective node expands the variable \texttt{host} (on l. \ref{l:tcpdump})
set to the other host's IP address (ll. \ref{l:varA},\ref{l:varB}).

The \texttt{synchronize} statement (l. \ref{l:sync-pings}) blocks until the
\texttt{doPing} tasklists have finished (l. \ref{l:ping}).  The
final step (l. \ref{l:getData}) copies the captured traffic from the monitor
node to the controller.

\section{User Studies}
\label{sec:usecase}

In the following section, we present an overview of projects using \gplmt  to
show the various different use cases and purposes \gplmt can be used for and
highlight the challenges emerging with respect to both experimentation as well
as using the \gplmt framework. Based on these experiences, we modified \gplmt
in the current version to cope with this challenges.

\subsection{The GNUnet Project - Large-scale Software Deployment in Heterogeneous Testbeds}

GNUnet\footnote{\url{https://gnunet.org}} is a GNU free software project
focusing on a future, decentralized Internet. GNUnet develops the GNUnet
peer-to-peer (P2P) framework to allow developers to realize  decentralized networking
applications.

GNUnet employs \gplmt to deploy the GNUnet framework to a large number of \pl
nodes to be able to test the software under real-world conditions and to support
bootstrapping of the network. GNUnet's requirement was to compile the latest
GNUnet version on \pl nodes directly.
GNUnet used \gplmt to provide the nodes with all software dependencies required.
While running, GNUnet was monitored to analyze the behavior of the software and
the P2P network and to obtain log files in in case of a crash. With \gplmt
detailed information for every node could be obtained.

For GNUnet, the major challenge was the unreliability and heterogeneity of the
\pl testbed. With a large number of nodes only a fraction were accessible and
working correctly. \pl nodes only provide outdated software and are very
heterogeneous both with respect to versions of the operating system and version
of software installed. Nodes also often get unavailable during operation.

\subsection{OpenLab Eclectic - A Holistic Development Life Cycle for P2P Applications}

The OpenLab Eclectic
Project\footnote{\url{http://www.ict-openlab.eu/experiments-use-cases/experiments.html}}
focused on developing a holistic development life cycle for distributed
systems by closing the gap between the testbed and the P2P community.

Eclectic used \gplmt to orchestrate, control and monitor networking, P2P
testing, and experimentation on different testbeds. \gplmt's functionality to
define experiments and to interact with testbeds using an abstraction layer
allowed Eclectic to deploy distributed systems on local systems, HPC's systems
like the SuperMUC\footnote{\url{https://www.lrz.de/services/compute/supermuc/}}
and Internet testbeds like \pl.

The main challenge for Eclectic was to define testbed independent experiments to
be able to transfer experiments between different testbeds. \gplmt was also used to
setup network nodes and collect experimental results. Within this project, \gplmt
was integrated with the Zabbix\footnote{\url{http://www.zabbix.com/}} network
monitoring solution to provide an integrated approach for infrastructure
monitoring and experiment scheduling.

\subsection{Testbed Management for Attack \& Defense Scenarios}

Datasets to train and test Intrusion Detection Systems (IDS) under realistic and
reproducible conditions are hard to obtain and generate. Such datasets have to
provide a high diversity of attacks with a high packet frequency but also have
ensure reproducible results and provide a clear labeled information about the
data flows.

At TUM's chair for network architectures and services, researchers used \gplmt to generate such datasets with different attack scenarios. To
generate such datasets, a virtualized testbed environment with virtual machines
grouped into attackers, victims and monitoring machines was used. These machines
were used to execute attacks as well as provide defense mechanisms and obtain
the generated network traffic. In addition, this testbed was used to evaluate
the quality of port scanners and port scan detection tools with the results
being collected and interpreted afterwards.

The main challenge was the grouping of the different entities, as
well as the complex interaction and nesting of tasks assigned to the entities.
Timing aspects as well as synchronization were crucial to this setting. The
monitoring and generation of test datasets during the experiment executions was
an additional challenge to be mastered.

\subsection{Distributed Internet Security Analysis}

In~\cite{2015jaros}, security researchers developed a distributed, \pl-based
approach to conduct large-scale scans of today's TLS deployment in the wild.
They used \pl nodes to perform distributed scans of large IP ranges and analyzed
the TLS certificates found on hosts. To conduct these scans, \gplmt was used to
deploy the scanning tool used to the \pl nodes, orchestrate the measurements,
and obtain results from the nodes.

A major challenge in this use case was long lasting scan experiments in
combination with the large number of parallel SSH connections established to \pl
nodes. The organization's intrusion detection system detected these connections
as a malicious attack and blocked the control node as the source of these
connections on the network as a consequence.

The main challenge was the large number of connections to the \pl nodes.
First, those connections had to be throttled during the experiment.
Apart from this, the number of connections established had to be managed.


\section{Future Work}

For future versions, we plan to decouple the \gplmt controller from the experimenter's host and
instead run \gplmt as a service on a dedicated control node. Users would then
submit experiments to the \emph{experiment queue} of a testbed, which is
managed by \gplmt. This would ease the use of shared testbeds.
Future versions of \gplmt may support target types other than SSH and \pl, for
example \emph{mobile devices}.
An intuitive user interface would ease experiment monitoring and control. This
feature was provided based on Zabbix in an earlier version of \gplmt but is not
available at the moment due to a recent refactoring of the code base.
\section{Conclusion}

The focus of \gplmt is to provide a lightweight and convenient way for
experimenters to conduct network experiments and manage testbed environments.
Instead of using handcrafted onetime scripts for every experiment, we envision
\gplmt to be flexible tool usable for different scenarios and use cases.
Using a high-level description language \gplmt offers opportunities to share
experiment descriptions among researchers and supports closer collaborations between experimenters.
Moreover, \gplmt's language was designed to support error handling, nested execution flows
and different timing aspects to provide a high level flexibility and adoptability.
\gplmt is still under active development and will be extended in the future.
With this work, we want to present \gplmt to the community and make it available for a broad audience.
\gplmt is free software and can be obtained from the
repository\footnote{\url{https://github.com/docmalloc/gplmt}}.

\textbf{Acknowledgments} 
This work has been supported by the German Federal Ministry of Education and Research (BMBF) under support code 16KIS0145, project SURF.
The authors would like to thank Matthias Jaros, Oliver Gasser for their helpful feedback, Omar Tarabai for his work on \gplmt and the integration
with Zabbix.

\bibliography{ref}
\bibliographystyle{splncs}

\end{document}